\documentclass[aps,prd,twocolumn,superscriptaddress,showpacs,amsmath,amssymb,amsthm,nofootinbib, preprintnumbers]{revtex4}
 \usepackage{amsmath}
\usepackage{graphicx}
\usepackage{epstopdf}
\usepackage{float}
\usepackage{hyperref}
\usepackage{color}
\usepackage[T1]{fontenc}
\usepackage[utf8]{inputenc}
\usepackage[toc,page]{appendix}
\usepackage[usenames,dvipsnames]{xcolor}
\usepackage[normalem]{ulem}

\newcommand{\be}{\begin{equation}}
\newcommand{\ee}{\end{equation}}
\newcommand{\ba}{\begin{eqnarray}}
\newcommand{\ea}{\end{eqnarray}}

\newcommand{\beq}{\begin{equation}}
\newcommand{\eeq}{\end{equation}}
\newcommand{\beqa}{\begin{eqnarray}}
\newcommand{\eeqa}{\end{eqnarray}}

\begin{document}

%%%%%%%%%%%%%%%%%%%%%%%%%%%%%%%%%%%%%%%%%%%%%%%%%%%%%%%%%%%%%%%%%%%%%%%%%%%%%%%%%%%%%%%%%%

%\title{Accelerated black holes in ModMax theory}
%\title{Accelerated Black Holes in Nonlinear Electrodynamics}
%\title{Accelerated Black Holes in ModMax Theory}
\title{Accelerated Black Holes beyond Maxwell's Electrodynamics}

\author{Jos{\'e} Barrientos}
\email{barrientos@math.cas.cz}
\affiliation{Institute of Mathematics of the Czech Academy of Sciences, {\v Z}itn{\'a} 25, 11567 Praha 1, Czech Republic}
\affiliation{Departamento de Enseñanza de las Ciencias B\'asicas, Universidad Cat\'olica del Norte, Larrondo 1281, Coquimbo, Chile}

\author{Adolfo Cisterna}
\email{adolfo.cisterna.r@mail.pucv.cl}
\affiliation{Sede Esmeralda, Universidad de Tarapac{\'a}, Av. Luis Emilio Recabarren 2477, Iquique, Chile}

\author{David Kubiz\v n\'ak}
\email{david.kubiznak@matfyz.cuni.cz}
\affiliation{Institute of Theoretical Physics, Faculty of Mathematics and Physics,
Charles University, Prague, V Hole{\v s}ovi{\v c}k{\' a}ch 2, 180 00 Prague 8, Czech Republic}

\author{Julio Oliva}
\email{julioolivazapata@gmail.com}
\affiliation{Departamento de F{\'i}sica, Universidad de Concepci{\'o}n, Casilla, 160-C, Concepci{\'o}n, Chile}

%\date{\today}

\date{May 31, 2022}

\begin{abstract}
%\tcr{\bf PLEASE PUT YOU CORRECT AFFILIATIONS AND EMAILS} 
Theories of non-linear electrodynamics naturally describe deviations from Maxwell's theory in the strong field regime. 
%electromagnetic fields. 
Among these, of special interest is the recently discovered ModMax electrodynamics, which is a unique 1-parametric generalization of Maxwell's theory that possesses the conformal invariance as well as the electromagnetic duality. In this paper we construct the asymptotically AdS accelerated black holes in this theory and study their thermodynamics -- providing thus a first example of accelerated solutions coupled to non-linear electrodynamics. Our study opens a window towards studying radiative spacetimes in non-linear electromagnetic regime as well as raises new challenges for their corresponding holographic interpretation.

\end{abstract}

\maketitle

%----------------------------------------------------------------------------------------%
\section{Introduction}
%----------------------------------------------------------------------------------------%

{
{\em Maxwell's electrodynamics} is one of the most remarkable and experimentally verified classical field theories ever constructed. Yet, it is theoretically plausible that %deviations from the linear Maxwell's theory  
even at the classical level
modifications of Maxwell's equations will occur in the strong field regime. The corresponding theories, that approach Maxwell in the weak field limit but modify the field equations in the vicinity of charges, are known as the theories of {\em Non-Linear Electrodynamics} (NLE). {Such theories can, for example, be utilized to provide a physical source for regular black holes \cite{Ayon-Beato:2000mjt}, as well as find their applications in more general contexts -- see the recent modern review \cite{Sorokin:2021tge}.}
}

{Perhaps the best known example of NLE is the {\em Born--Infeld theory} \cite{Born:1934gh}. It features remarkable mathematical properties, has unique propagation of waves,  and naturally appears in the context of string theory and early Universe cosmology. However, Born--Infeld electrodynamics is less symmetric than the Maxwell theory. While it possesses a fundamental {\em $SO(2)$ duality} \cite{Gibbons:1995cv}, it features a dimensionfull parameter and thence it is no longer  {\em conformally invariant}. For a long time it was thought that Maxwell's theory is the only theory with the above two fundamental symmetries. This is, however, not true. Not long ago, a 1-parametric generalization of the Maxwell theory with both these symmetries have been constructed. The corresponding theory is known as the {\em ModMax theory} \cite{Bandos:2020jsw, Kosyakov:2020wxv}. {(See also \cite{Bandos:2020hgy, Bandos:2021rqy, Kruglov:2021bhs, Nastase:2021uvc, Kuzenko:2021qcx, Avetisyan:2021heg, Cano:2021tfs, Babaei-Aghbolagh:2022uij} for recent extensions of this theory).}
}

{In order to better understand the effects that arise from non-linearities of the ModMax theory, it is useful to find some of its exact solutions. Interestingly, there is already a small catalogue of such solutions. Namely, it was shown that the spherically symmetric  
\cite{Flores-Alfonso:2020euz, BallonBordo:2020jtw}
and the Taub-NUT \cite{BallonBordo:2020jtw, Flores-Alfonso:2020nnd} solutions in the ModMax theory are (although slightly different from those of Maxwell) remarkably %similar to those of Maxwell 
`Maxwell-like', though this is no longer true in the presence of slow rotation \cite{Kubiznak:2022vft}. 
}

{
In this paper, we will focus on finding fields of uniformly {\em accelerated charges} and fields of self-gravitating {\em accelerated black holes} in the ModMax theory. Such solutions are of particular interest as they allow one to study electromagnetic and gravitational radiation, e.g. \cite{Bicak:2002yk, Podolsky:2003gm}, as well as are very interesting from the holographic point of view, e.g.  \cite{Anabalon:2018ydc, Anabalon:2018qfv}. In particular, we shall construct the ModMax generalization of the slowly accelerating charged AdS C-metric \cite{Griffiths:2005qp} and study its thermodynamics.  As we shall see, also this solution is remarkably similar to that of the 
Maxwell theory. As far as we know, these solutions provide a first example of charged accelerating black holes {\em beyond} Maxwell's theory. 
}

\section{Theories of NLE and the ModMax theory}
{Let us start by reviewing the general framework of NLE theories. 
To this purpose we consider the following Einstein-NLE action:}
%In what follows we are interested in solutions of the following action: 
\begin{equation}\label{bulkAct}
    I= \frac{1}{16\pi} \int_{M} d^4x \sqrt{-g}\left(R +\frac{6}{\ell^2}-4{\cal L}\right)\,,
\end{equation}
where $R$ stands for the Ricci scalar, $\ell$ is the AdS radius, related to the cosmological constant $\Lambda=-3/\ell^2$, and ${\cal L}$ is the Lagrangian of the corresponding NLE theory. In general this may depend on two electromagnetic invariants, 
\be\label{SP}
{\cal S}=\frac{1}{2}F_{\mu\nu} F^{\mu\nu}\,,\quad {\cal P}=\frac{1}{2}F_{\mu\nu} (*F)^{\mu\nu}\,, 
\ee
where the field strength $F_{\mu\nu}$ is given in terms of the 
vector potential $A_\mu$ by the familiar expression, 
$F_{\mu\nu}=\partial_\mu A_\nu-\partial_\nu A_\mu$, and 
$(*F)_{\mu\nu}=(1/2) \epsilon_{\mu\nu}{}^{\rho\lambda}F_{\rho\lambda}\,.$
 {Whereas 
${\cal S}$ is a true scalar, the invariant ${\cal P}$ is only a pseudoscalar. To restore parity invariance, we thus consider theories that depend on ${\cal P}$ via its `square':  
\be\label{general}
{\cal L}={\cal L}({\cal S},{\cal P}^2)\,. 
\ee

The generalized Einstein-NLE equations write as 
\be\label{FE}
 G_{\mu\nu}=8\pi T_{\mu\nu}\,,\quad 
d*E=0\,,\quad
%\nabla_\mu E^{\mu\nu}=0\,,\quad 
dF=0\,, 
\ee
where $E$ is a non-linear function of $F_{\mu\nu}$ and $(*F)_{\mu\nu}$, $E=E(F,*F)$, and $T_{\mu\nu}$ has a similar structure to that of Maxwell's theory. Namely, we have } 
\ba
8\pi T^{\mu\nu}&=&4F^{\mu\sigma}F^{\nu}{}_\sigma {\cal L_S}+2({\cal P}{\cal L_P}-{\cal L})g^{\mu\nu}\,,\label{Tmunu}\\
E_{\mu\nu} &=& \frac{\partial \mathcal{L}}{\partial F^{\mu\nu}}
=2\Bigl({\cal L_S}F_{\mu\nu}+{\cal L_P}*\!F_{\mu\nu}\Bigr)\,, \label{Emunu}
\ea
using the following notation:
\be
{\cal L}_{\cal S}=\frac{\partial {\cal L}}{\partial {\cal S}}\,,\quad  
{\cal L}_{\cal P}=\frac{\partial {\cal L}}{\partial {\cal P}}\,.
\ee
In what follows we shall calculate the 
electric and magnetic charges inside a closed spacelike two-surface $\mathcal{H}$. These are simply given by 
%\tcb{should we argue why these are the correct choices?} \textcolor{lime}{I think the line after them is enough for that, with the duality mentioned above.}
\be\label{charges}
Q_e = \frac{1}{4\pi} \int_{\mathcal{H}}* E\,, \quad Q_m = \frac{1}{4\pi}\int_{\mathcal{H}} F\,.
\ee

In particular, we will be interested in the following ModMax theory \cite{Bandos:2020jsw, Kosyakov:2020wxv}:
\be\label{NLELag}
    \mathcal{L} = \frac{1}{2} \left({\cal S}\cosh\gamma - \sqrt{{\cal S}^2+{\cal P}^2}\sinh\gamma\right)\,,
\end{equation}
a 1-parametric generalization of the Maxwell theory characterized by the dimensionless parameter $\gamma$. Setting $\gamma=0$ recovers the Maxwell's case. When $\gamma\neq 0$ a birefringence phenomenon occurs \cite{Bandos:2020jsw}: 
apart from the light-like polarization mode there exists another mode which is subluminal for $\gamma>0$ and superluminal for $\gamma<0$, hinting on a physical restriction $\gamma\geq 0$.  
The ModMax theory is distinguished by being both: invariant under conformal transformations of the metric, $g\to \Omega^2 g$, and invariant under the $SO(2)$ duality rotations 
\begin{equation}\label{dualityRot}
    \begin{pmatrix}
    E'_{\mu\nu}\\
     *F'_{\mu\nu}
    \end{pmatrix}
    =
    \begin{pmatrix}
        \cos\theta & \sin \theta\\
        -\sin\theta & \cos\theta 
    \end{pmatrix}
    \begin{pmatrix}
        E_{\mu\nu}\\
        *F_{\mu\nu}
    \end{pmatrix}\,,
\end{equation}
{where from \eqref{Emunu} we have 
\be\label{EE}
E=\Bigl(\cosh \gamma -\frac{{\cal S}\sinh\gamma}{\sqrt{{\cal S}^2+{\cal P}^2}}\Bigr)F-\frac{{\cal P}\sinh\gamma}{\sqrt{{\cal S}^2+{\cal P}^2}}*F\,.  
\ee
}

{
Obviously, solutions of the ModMax theory with vanishing invariant ${\cal P}$ coincide with those of the Maxwell theory. This is for example the case of static electrically/magnetically charged solutions. However, 
when both electric and magnetic charges are present, already static spherical solutions feature `small' deviations from the corresponding solutions of the Maxwell theory,   e.g. \cite{Flores-Alfonso:2020euz, BallonBordo:2020jtw} (see also \cite{BallonBordo:2020jtw, Flores-Alfonso:2020nnd} for the case of Taub-NUT solutions). On the other hand, the presence of slow rotation  reveals the full non-linearity of the theory \cite{Kubiznak:2022vft} and the full rotating solution is currently unknown. In what follows we concentrate on a particularly interesting class of solutions of the ModMax theory \eqref{NLELag} that describes  accelerated charges and black holes. }

\section{Accelerated charge in flat space}
The uniformly accelerated charges have puzzled researchers for many years. In Maxwell's theory, the corresponding electromagnetic field can be obtained with the help of the Lienard--Wiechert potentials, {e.g. \cite{eriksen2000electrodynamics}. 
This field is quite complicated in the Minkowski frame where it features time dependent electric and magnetic fields. However,  
when transformed to the coordinate system associated with an observer `sitting on a charge' (Rindler frame), the corresponding field significantly simplifies and becomes `static' --  `Coulomb-like'. }

{
For this reason, let us consider the accelerated Rindler frame  
 $(T, X, \rho, \varphi$),  }
\be\label{Rindler}
ds^2=-(1+AX)^2dT^2+dX^2+d\rho^2+\rho^2 d\varphi^2\,, 
\ee
related to the Minkowski cylindrical coordinates $(t_{\mbox{\tiny M}},x_{\mbox{\tiny M}},\rho,\varphi)$, by $t_{\mbox{\tiny M}}=(X+1/A)\sinh(AT)\,,x_{\mbox{\tiny M}}=(X+1/A)\cosh(AT)$, with $A$ the magnitude of the 4-acceleration, and $x$ the direction of motion of the charge. The corresponding Coulomb's vector potential is then modified due to the acceleration as follows:
\be
B= -\frac{Q_e(2+Ay)}{r}dT+\frac{Q_m y}{r}d\varphi\,,
\ee   
where 
\be
y=2X+A(X^2+\rho^2)\,,\quad r=\sqrt{y^2+4\rho^2}\,. 
\ee
One can easily check that it satisfies the Maxwell equations in the Rindler frame \eqref{Rindler}.

{Finding the corresponding solution for any NLE different from Maxwell is (due to the lack of  generalized Lienard--Wiechert potentials) a surprisingly complicated task. However,  one can easily check that (up to the rescaling of electric and magnetic charges), the above solution remains valid also for the ModMax electrodynamics. Namely, we find that the following potential: 
\be
B= -\frac{Q_ee^{-\gamma}(2+Ay)}{r}dT+\frac{Q_m y}{r}d\varphi\,
\ee   
satisfies all the ModMax equations in the Rindler frame \eqref{Rindler}. Note that due to the presence of both $Q_e$ and $Q_m$, this solution is non-trivial as the invariant does not vanish and reads ${\cal P}=-32Q_eQ_me^{-\gamma}/r^4$.}
As we shall see in the next section, this `test field solution' can be extended to the full solution of the ModMax--Einstein system, where the test accelerated charge is `replaced' with the self-gravitating accelerated black hole.

\section{Accelerated black holes beyond Maxwell}

\subsection{Solution}
{Let us now present the full self gravitating solution of the  Einstein--ModMax theory describing a slowly accelerated black hole in AdS space. The solution generalizes the (Maxwell) charged AdS C-metric, e.g. \cite{Griffiths:2005qp}. Employing $(t,r, \theta, \varphi)$-type coordinates used in   \cite{Appels:2016uha, Anabalon:2018ydc, Anabalon:2018qfv}, the corresponding generalized AdS C-metric reads:}
\be
ds^2=\frac{1}{\Omega^2}\Bigl(-\frac{fdt^2}{\alpha^2}+\frac{dr^2}{f}+r^2\Bigr[\frac{d\theta^2}{h}+h\sin^2\!\theta \frac{d\varphi^2}{K^2}\Bigr]\Bigr)\,, 
\ee
where the metric functions $f, h$ and $\Omega$ take the following form:
\ba 
f&=&(1-A^2r^2)f_0+\frac{r^2}{\ell^2}\,,\nonumber\\
h&=&1+2mA\cos\theta+A^2z^2\cos^2\!\theta\,,\nonumber\\
\Omega&=&1+Ar\cos\theta\,,\quad 
z^2=e^{-\gamma}(q_e^2+q_m^2)\,,
\ea
and 
\be
f_0=1-\frac{2m}{r}+\frac{z^2}{r^2}\,,
\ee 
which would be the static metric function characterizing the (asymptotically flat) static solution. The angular variable $\varphi$ has periodicity $2\pi$, and the time coordinate $t$ was rescaled by a constant $\alpha$ (see below). The metric is accompanied with the following ModMax field:
\be
F=dB,\quad B=-\frac{q_ee^{-\gamma}}{\alpha r}dt+q_m\cos\theta \frac{d\varphi}{K}\,, 
\ee
and is {characterized by the following two invariants:
\be
 {\cal S}=\frac{\Omega^4\bigl(q_m^2-(q_e e^{-\gamma})^2\bigr)}{r^4}\,,\quad 
 {\cal P}=\frac{2\Omega^4 q_e e^{-\gamma}q_m}{r^4}\,.
\ee
It yields an exact solution to the Einstein--ModMax equations.
}

{
This solution is characterized by the following six independent parameters: $\{m, q_e, q_m,  \ell, A, K\}$. These are related to the properties of the black hole, such as its mass, electric and magnetic charges, AdS radius, black hole acceleration, and conical deficits in the spacetime. Alternatively, we may re-express the two parameters $A$ and $K$ in terms of the conical deficits of the north-pole ($\delta_+$) and south-pole ($\delta_-$) axes, or more physically, in terms of the corresponding cosmic string tensions   }
\be
\mu_\pm=\frac{\delta_\pm}{8\pi}=\frac{1}{4}\Bigl[1-\frac{\Xi\pm 2mA}{K}\Bigr]\,,
\ee 
where
\be
\Xi=1+A^2z^2\,. 
\ee 
{
In this physical picture, the black hole is pulled by two cosmic strings situated on the symmetry axis, and the difference between their tensions causes the black hole to accelerate. The constant $\alpha$ is not an independent parameter -- similar to the Maxwell case \cite{Anabalon:2018qfv}, we define it to be 
\be\label{alpha}
\alpha=\sqrt{\Xi(1-A^2\ell^2 \Xi)}\,. 
\ee
As we shall see, this choice yields  well defined  variational principle as well as makes thermodynamics consistent.  
}

{
Similar to the Maxwell case \cite{Anabalon:2018qfv}, we have to impose a number of restrictions on the parameters of the solution, in order the above metric is Lorentzian and describes a (spherical) slowly accelerating black hole. Namely,   i) For the standard interpretation of the coordinate $\theta$ on $[0, \pi]$, we must have $h(\theta)>0$, 
giving 
\be
mA<\begin{cases}
\frac{1}{2}\Xi\quad\mbox{for}\quad \Xi\in(0,2]\,,\\
\sqrt{\Xi-1}\quad  \mbox{for}\quad \Xi>2\,.
\end{cases}
\ee
ii) The conformal boundary is located at $\Omega=0$, that is, $r_{\mbox{\tiny b}}=-1/(A\cos\theta)$. On the other hand, for the black hole interpretation we need a horizon in the bulk -- located at $f(r_+)=0$. Thus $f(r)$ needs to have at least one root in the range $r\in (0,1/A)$.
iii) The black hole accelerates slowly when there is no acceleration horizon reaching all the way till infinity. We thus have to require that on the boundary
$f(r=-1/(A\cos\theta))$ has no roots. Similar to the Maxwell case, this condition already guarantees that the above defined $\alpha$ is  real, that is, $1 \geq A^2\ell^2\Xi$. 
%\tcr{\bf Is this obvious?}
}

\subsection{Thermodynamics}

{
Let us now turn towards thermodynamics of these black holes, following the strategy developed in \cite{Appels:2017xoe, Anabalon:2018ydc, Anabalon:2018qfv} {(see also \cite{Astorino:2016ybm, Cassani:2021dwa, Ball:2020vzo}). Namely,} we want to identify thermodynamic quantities obeying the following first law:
\begin{equation}\label{firstLaw}
\delta M=T\delta S+\phi_e \delta Q_e+\phi_m \delta Q_m+V\delta P-\lambda_+\delta \mu_+-\lambda_-\delta \mu_-\,,  
\end{equation}
together with the corresponding Smarr relation
\begin{equation}\label{Smarr}
M=2TS-2PV+\phi_eQ_e+\phi_m Q_m\,.
\end{equation}
Here, apart from the standard black hole charges and their conjugates, we observe the $+V\delta P$ term, where $P$ is the thermodynamic pressure induced by the negative cosmological constant, 
\be
P=-\frac{\Lambda}{8\pi}=\frac{3}{8\pi \ell^2}\,, 
\ee
and $V$ is its conjugate quantity that was properly introduced into the black hole thermodynamics in \cite{Kastor:2009wy}. We also observe two extra work terms, $-\lambda_\pm \delta \mu_\pm$, due to the presence of cosmic strings. Namely, $\lambda_\pm$ are the conjugates to string tensions $\mu_\pm$ that are  called the thermodynamic lengths \cite{Appels:2017xoe}. 
The Smarr relation \eqref{Smarr} is consistent with the first law \eqref{firstLaw} via the dimensional scaling argument \cite{Kastor:2009wy, Appels:2017xoe}.
}

Let us now proceed and identify all the remaining thermodynamic quantities.  As always, the simplest to calculate  
are the black hole temperature and its entropy (given by the Bekenstein's area law). We  have 
\be\label{S}
S=\frac{\mbox{Area}}{4}=\frac{r^2}{4K}\int \frac{\sin\theta}{\Omega^2}d\theta d\varphi\Bigr|_{r=r_+}=\frac{\pi r_+^2}{K(1-A^2r_+^2)}\,, 
\ee
and
\ba
T&=&\frac{f'(r_+)}{4\pi \alpha}\\
&=&\frac{1+\frac{3r^2_+}{l^2}-A^2r_+^2\bigl(2+\frac{r_+^2}{l^2}-A^2r_+^2\bigr)}{4\pi\alpha r_+(1-A^2r_+^2)}-\frac{z^2(1-A^2r_+^2)}{4\pi\alpha r_+^3}\,.\nonumber
\ea
Next, using the definition of the dual tensor $E_{\mu\nu}$, \eqref{EE}, and formulae \eqref{charges}, we find the following electric and magnetic charges:
\ba
Q_e&=&\frac{1}{4\pi}\int *E=\frac{q_e}{K} \,,\\
Q_m&=&\frac{1}{4\pi}\int F=\frac{q_m}{K}\,,
\ea
reflecting the electromagnetic duality 
\be\label{EMduality}
q_m \quad \leftrightarrow \quad q_e\,. 
\ee
The mass of the black hole is also relatively easy to calculate, for example by employing the conformal methods \cite{Ashtekar:1999jx}. Namely, using the Killing vector $\partial_t$,  we find  
\be
M=\frac{m}{\alpha K}(1-A^2\ell^2 \Xi)\,. 
\ee

{
A bit more tricky is to calculate the electrostatic potential. Naively, one may identify $\phi_e(r)=-(\partial_t)\cdot B$. When evaluated on the horizon  ($r=r_+)$ this yields 
\be\label{phie}
\phi_e=\frac{q_e e^{-\gamma}}{\alpha r_+}\,. 
\ee 
However, the corresponding quantity  on the boundary does not vanish  and is (due to the presence of acceleration) $\theta$-dependent. Thus, similar to the Maxwell case \cite{Anabalon:2018qfv}, in order to justify that \eqref{phie} is the correct thermodynamic potential, one needs to use the (properly generalized) Hawking--Ross prescription \cite{Hawking:1995ap}. This goes as follows. One considers a Wick-rotated Euclidean version of the metric and gauge potential, while transforming to a gauge where $\phi_e(r)$ vanishes on the horizon. 
The thermodynamic potential is then defined via the (generalized to NLE) Hawking--Ross term 
%\tcr{\bf do we want to provide the derivation of this in appendix? I would rather leave it to the bigger paper.}
\begin{equation}\label{HR}
\phi_e=\frac{1}{4\pi Q_e \beta}\int_{\partial M}\sqrt{h}n_\mu E^{\mu \nu}A_\nu\,,
\end{equation}
where $E_{\mu\nu}$ is the dual tensor \eqref{Emunu}, $\beta=1/T$ is the inverse periodicity in Euclidean time, and $n^\mu$ is the  outward pointing unit normal to $\partial M$ (which in our case is given by $\Omega=0$). One can easily confirm that this indeed yields \eqref{phie}. Moreover, employing the electromagnetic duality \eqref{EMduality} immediately yields the magnetic potential 
\be\label{phim}
\phi_m=\frac{e^{-\gamma}q_m}{\alpha r_+}\,.
\ee
}

{Finally, the remaining conjugate thermodynamic quantities are given by 
\ba
V&=&\Bigl(\frac{\partial M}{\partial P}\Bigr)_{S,Q_e,\dots}\!\!=\frac{4\pi}{3K\alpha}\left[\frac{r_+^3}{(1-A^2r_+^2)^2}+mA^2l^4\Xi\right]\,,\nonumber\\
\lambda_{\pm}&=&
-\Bigl(\frac{\partial M}{\partial \mu_\pm}\Bigr)_{S,Q_e,\dots}\!\!=
\frac{r_+}{\alpha(1\pm Ar_+)}-\frac{m}{\alpha\Xi}\mp\frac{Al^2\Xi}{\alpha}\,,\nonumber\\
\ea  
where we have used the explicit expression \eqref{alpha} for $\alpha$. 
It is then easy to verify that with these both the generalized first law \eqref{firstLaw} and the Smarr relation \eqref{Smarr} are satisfied. 
}

{An `independent' check of the above thermodynamic quantities is provided by the Euclidean action calculation. Namely, we identify  the Gibbs free energy $G$ with the (total) Euclidean action, $G=I_{\mbox{\tiny T}}/\beta$, where the bulk action $I$, \eqref{bulkAct}, is supplemented by the York--Gibbons--Hawking term and the AdS counterterms:
\be
I_{\mbox{\tiny T}}=I 
+\frac{1}{8\pi}\int_{\partial M} \sqrt{h}\left[\mathcal{K}-\frac{2}{l}-\frac{l}{2}\mathcal{R}(h)\right]\,,
\ee 
where $\mathcal{K}$ and $\mathcal{R}(h)$ are the extrinsic and Ricci scalar of the conformal boundary, respectively. From here we obtain }
\ba
G&=&\frac{m(1-2A^2 l^2\Xi)}{2\alpha K}-\frac{r^3_+}{2K\alpha l^2(1-A^2r_+^2)^2}\nonumber\\
&&-\frac{e^{-\gamma}(q_e^2-q_m^2)}{2K\alpha r_+}\,.
\ea
One can easily check that the obtained $G$ then satisfies
\be
G=G(T,\phi_e, Q_m, \mu_+,\mu_-, P)=M-TS-\phi_e Q_e\,, 
\ee
as expected \cite{Caldarelli:1999xj}. One can then `independently arrive' at various thermodynamic quantities. For example, the magnetic potential and entropy are then given by
\be
\phi_m=\Bigl(\frac{\partial G}{\partial Q_m}\Bigr)_{T,\phi_e, \mu_\pm, P}\,\quad 
S=-\Bigl(\frac{\partial G}{\partial T}\Bigr)_{\phi_e,Q_m, \mu_\pm, P}\,,
\ee
and are easily verified to yield \eqref{S} and \eqref{phim}, respectively.  

{
Having constructed the consistent thermodynamics of the above solution, the next step would be to study the holographic interpretation of the constructed solution, and in particular to analyze the holographic stress energy tensor $\tau_{ab}$. Interestingly, as noticed in \cite{Anabalon:2018ydc, Anabalon:2018qfv} (see also \cite{Cassani:2021dwa, Wang:2022hzh}), holography can also be used to determine the expression for $\alpha$. Namely, one demands that (for fixed $P$ and $\mu_\pm$)
\be
\delta I_{\mbox{\tiny T}}=-\frac{1}{2}\int_{\partial M}\tau_{ab}\delta h^{ab}\sqrt{-h}d^3x=0\,, 
\ee 
which yields the expression  
\eqref{alpha}. We shall leave the details of these calculations to \cite{usprep}.
}

\section{Summary}
{
To summarize, in this paper we have constructed the first example of fields of accelerated charges and black holes beyond the Maxwell electrodynamics.  Namely, we have constructed a field of the uniformly accelerated ModMax charge, as well as the 1-parametric generalization of the charged AdS C-metric in the ModMax theory. We have seen that such solutions are  remarkably `Maxwell-like'.
% and so is the associated black hole thermodynamics. 
This similarity %with Maxwell 
stems from the `static character' of these solutions --  the observed small deviations from Maxwell can be traced to the interaction of electric and magnetic charges (giving rise to non-trivial invariant ${\cal P}$). 
}

{In order to formulate properly the associated black hole thermodynamics, 
we have generalized the Hawking--Ross prescription for the electric potential, which for any NLE theory  is given by formula \eqref{HR}. The resultant thermodynamics is also remarkably similar to the Maxwell case. The holographic properties of the newly constructed solutions will be presented elsewhere \cite{usprep}.}

{
As a next step, it would be interesting to probe the effect of non-linearities of the electromagnetic field on the bulk \cite{Abbasvandi:2018vsh, Abbasvandi:2019vfz} and boundary  phase transitions. The latter is likely to generalize the findings of \cite{Dutta:2013dca, Cong:2021jgb} to the accelerating and ModMax charged situations. It would also be interesting to find accelerated black holes in other NLE theories.   
}

%\tcr{\bf I think we should not mention here the conformal scalar -- just to give us more time to write it up in the longer paper :)}

\section*{Acknowledgements}
%We would like to thank the anonymous referee for helping us to improve our manuscript.  
{We would like to thank Ji{\v r}{\'i} Bi{\v c}{\'a}k for motivating one of the authors (D.K.) to search for  accelerating solutions in NLE -- a task that took almost 20 years to complete.}
The work of J.B. is supported by the  ``Programme to support prospective human resources – post Ph.D. candidates''  of the Czech Academy of Sciences, project L100192101.
A.C work is funded by FONDECYT Regular grant No. 1210500, Beca Chile de Postdoctorado Grant No. 74200012 and PROGRAMA DE COOPERACI\'{O}N CIENT\'{I}FICA ECOSud-CONICYT 180011/C18U04.
D.K. acknowledges the partial support  from %the Perimeter Institute for Theoretical Physics and the
%D.K
the Natural Sciences and Engineering Research Council of Canada (NSERC). J.O. is funded by FONDECYT Regular grant No. 1221504 and Proyecto de Cooperaci\'on Internacional 2019/13231-7 FAPESP/ANID.

%T.T was supported by Research Grant No. GA\v{C}R 21-11268S and O.S by Research Grant No. GA\v{C}R 22-14791S.
%Research at Perimeter Institute is supported in part by the Government of Canada through the Department of Innovation, Science and Economic Development Canada and by the Province of Ontario through the Ministry of Colleges and Universities. 
%Perimeter Institute and the University of Waterloo are situated on the Haldimand Tract, land that was promised to the Haudenosaunee of the Six Nations of the Grand River, and is within the territory of the Neutral, Anishnawbe, and Haudenosaunee peoples.

%\bibliography{references}
%\bibliographystyle{JHEP}

\providecommand{\href}[2]{#2}\begingroup\raggedright\endgroup

\end{document}